\pdfoutput=1
\documentclass[aps,prl,twocolumn,noshowpacs,a4paper]{revtex4}

\usepackage{graphicx}
\usepackage{color}
\usepackage{textcomp}
\usepackage{amssymb}\usepackage{amsmath}\usepackage{amsthm}

\begin{document}

\title{Cavity Opto-Mechanics with a Bose-Einstein Condensate}

\author{Ferdinand Brennecke$^\dagger$}
\author{Stephan Ritter$^\dagger$}
\author{Tobias Donner}
\author{Tilman Esslinger}
\email{esslinger@phys.ethz.ch}
\homepage{http://www.quantumoptics.ethz.ch}
\affiliation{$^\dagger$These authors contributed equally to this work.\\
Institute for Quantum Electronics, ETH Z\"{u}rich, CH--8093 Z\"{u}rich, Switzerland}

\date{\today}

\begin{abstract}
Cavity opto-mechanics studies the coupling between a mechanical oscillator and a cavity field, with the aim to shed light on the border between classical and quantum physics. Here we report on a cavity opto-mechanical system in which a collective density excitation of a Bose-Einstein condensate is shown to serve as the mechanical oscillator coupled to the cavity field. We observe that a few photons inside the ultrahigh-finesse cavity trigger a strongly driven back-action dynamics, in quantitative agreement with a cavity opto-mechanical model. With this experiment we approach the strong coupling regime of cavity opto-mechanics, where a single excitation of the mechanical oscillator significantly influences the cavity field. The work opens up new directions to investigate mechanical oscillators in the quantum regime and quantum gases with non-local coupling.
\end{abstract}

\maketitle

Cavity opto-mechanics has played a vital role in the conceptual exploration of the boundaries between classical and quantum-mechanical systems \cite{braginsky1980}. These fundamental questions have recently found renewed interest through the experimental progress with micro-engineered mechanical oscillators. Indeed, the demonstration of laser cooling of the mechanical mode \cite{hohberger2004a,schliesser2006,arcizet2006,gigan2006,corbitt2007,thompson2008} has been a substantial step towards the quantum regime \cite{mancini1997,marshall2003,zhang2003}.

In general, light affects the motional degrees of freedom of a mechanical system through the radiation pressure force, which is caused by the exchange of momentum between light and matter. In cavity opto-mechanics the radiation pressure induced interaction between a single mode of an optical cavity and a mechanical oscillator is investigated. This interaction is mediated by the optical path length of the cavity which depends on the displacement of the mechanical oscillator.

New possibilities for cavity opto-mechanics are now emerging in atomic physics by combining the tools of cavity quantum electrodynamics (QED) \cite{hood2000,pinkse2000} with those of ultracold gases. Placing an ensemble of atoms inside a high-finesse cavity dramatically enhances the atom-light interaction since the atoms collectively couple to the same light mode \cite{nagorny2003,black2003,slama2007,gupta2007,colombe2007,brennecke2007}. In the dispersive regime this promises an exceedingly large opto-mechanical coupling strength, tying the atomic motion to the evolution of the cavity field. Recently, a thermal gas prepared in a stack of nearly two-dimensional trapping potentials has been shown to couple to the cavity field by a collective center of mass mode leading to Kerr nonlinearity at low photon numbers \cite{gupta2007} and back-action heating induced by quantum-force fluctuations \cite{murch2008}.

\begin{figure}
\centering
\includegraphics{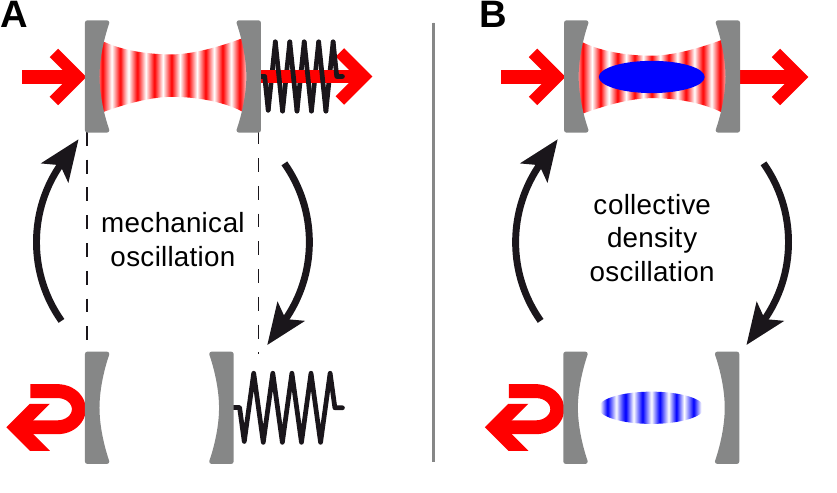}
\caption{(\textbf{A}) Cavity opto-mechanical model system. A mechanical oscillator, here one of the cavity mirrors, is coupled via radiation pressure to the field of a cavity whose length depends on the oscillator displacement. (\textbf{B}) Coupling a  Bose-Einstein condensate dispersively to the field of an optical high-finesse cavity constitutes an equivalent system. Here a collective density excitation of the condensate acts as the mechanical oscillator which strongly couples to the cavity field. Feedback on the cavity field is accomplished by the dependence of the optical path length on the atomic density distribution within the spatially periodic cavity mode structure. In contrast to opto-mechanical systems presented so far this mechanical oscillator is not based on the presence of an external harmonic potential (e.g.~a spring). It is rather provided by kinetic evolution of the condensate density excitation.}
\end{figure}

A crucial goal for cavity opto-mechanical systems is the preparation of the mechanical oscillator in its ground state with no thermally activated excitations present, yet at the same time providing strong coupling to the light field. Here we use a Bose-Einstein condensate as the ground state of a mechanical oscillator and thereby suppress thermal excitations of the oscillator to an unprecedented level. The cavity field couples to a collective density excitation of the Bose-Einstein condensate which matches the cavity mode, resulting in an exceedingly large coupling strength. Despite the absence of an external restoring force for the mechanical oscillator, the framework of cavity opto-mechanics can be applied since only a single excitation mode of the Bose-Einstein condensate is involved, see Fig.~1.

In our experimental setup \cite{brennecke2007,ottl2006} a Bose-Einstein condensate of typically $1.2\times 10^5$ $^{87}\mathrm{Rb}$ atoms in the $|F, m_F\rangle = |1, -1\rangle$ ground state is coupled to the field of an optical ultrahigh-finesse Fabry-Perot cavity. Our system is in the strong coupling regime of cavity QED, i.e.~the maximum coupling strength between a single atom and a single $\sigma^-$ polarized intracavity photon $g_0 = 2\pi \times 10.9$\,MHz is larger than both the amplitude decay rate of the atomic excited state $\gamma = 2\pi \times 3.0$\,MHz and that of the intracavity field $\kappa = 2\pi \times 1.3$\,MHz. Trapping the condensate within the cavity is accomplished by a crossed-beam dipole trap with trap frequencies $(\omega_x, \omega_y, \omega_z) = 2 \pi \times (222, 37, 210)$\,Hz, where $x$ denotes the cavity axis and $z$ the vertical axis. The cavity has a length of 178 \textmu m and its $\mathrm{TEM}_{00}$ mode has a waist of 25 \textmu m. The mode maximally overlaps with the condensate having Thomas-Fermi radii of $(R_x, R_y, R_z) = (3.3, 20.0, 3.5)$\textmu m. All experiments presented here were performed without active stabilization of the cavity length, which would give rise to an additional standing wave potential for the atoms \cite{gupta2007,murch2008,brennecke2007}.

The coupled dynamics of the Bose-Einstein condensate and the cavity field is driven by continuously applying a weak pump laser field along the cavity axis (see Fig.~1). The light transmitted through the cavity is monitored using a single-photon counter and serves as a probe for the dynamics of the system. With a detuning of $\Delta_a = \omega_p - \omega_a \geq 10^4\gamma$ between pump laser frequency $\omega_p$ and atomic $D_2$ line transition frequency $\omega_a$ spontaneous emission can be mostly neglected.

\begin{figure}
\centering
\includegraphics{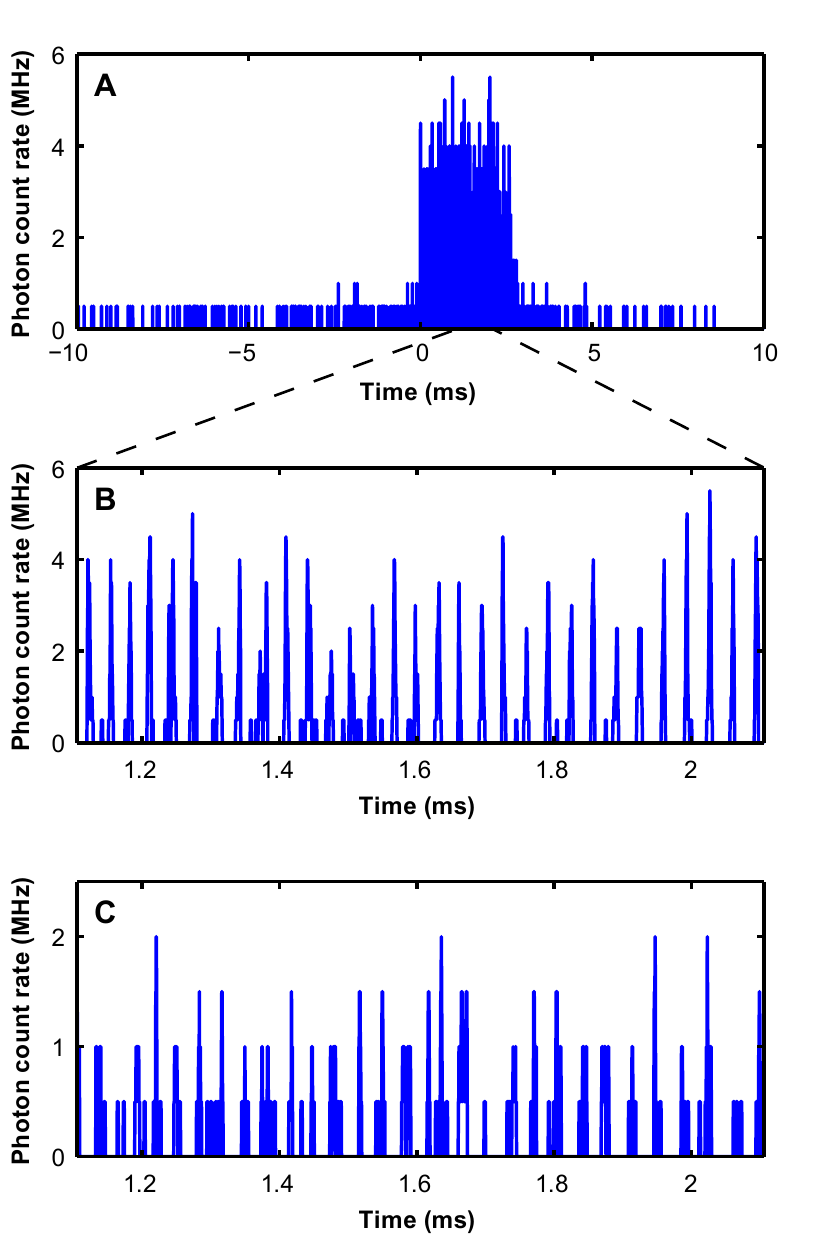}
\caption{(\textbf{A-C}) Response of the continuously driven BEC-cavity system. Shown is a single trace of the cavity transmission (averaged over 2\,\textmu s) while scanning the cavity-pump detuning at a rate of $+2\pi\times 2.9\,\mathrm{MHz}/\mathrm{ms}$ across its $\sigma^-$ resonance \cite{brennecke2007}. The pump rate corresponds to a mean intracavity photon number on resonance of $7.3\pm 1.8$ (A and detail B) and $1.5\pm 0.4$ (C). The photon count rate for one mean intracavity photon is $0.8\pm 0.2\,\mathrm{MHz}$. The dead time of the single-photon counter is $50\,\mathrm{ns}$ which leads to a saturation of high photon count rates. The pump laser was blue detuned by $\Delta_a = 2 \pi \times 32$\,GHz with respect to the atomic resonance.}
\end{figure}

Figure 2 shows the response of the system while scanning the pump frequency across the optical resonance. We observe a characteristic transmission signal (Fig.~2A) which exhibits a sharp rising edge and subsequently regular and fully modulated oscillations (Fig.~2B) lasting for about 2.5\,ms. These oscillations start at a frequency of about 37\,kHz which slightly decreases over the train of oscillations and does not depend on the speed at which the pump frequency is varied. Similar responses of the system were measured for lower pump strengths at the same detuning (Fig.~2C) as well as for pump-atom detunings of up to $\Delta_a=2\pi\times 300\, \mathrm{GHz}$, provided the pump rate was increased sufficiently. Moreover, when continuing the pump frequency scan we observe a second train of oscillations in the vicinity of the stronger coupling $\sigma^+$ resonance \cite{brennecke2007}. This is in accordance with the observation that the condensate remains intact during probing, which is directly inferred from absorption images taken subsequent to probing. The observed oscillatory behavior is obviously in strong contrast to a Lorentzian shaped resonance curve which would be expected for an atomic ensemble frozen inside the cavity, i.e.~when the atomic external degree of freedom is neglected.

\begin{figure*}
\centering
\includegraphics{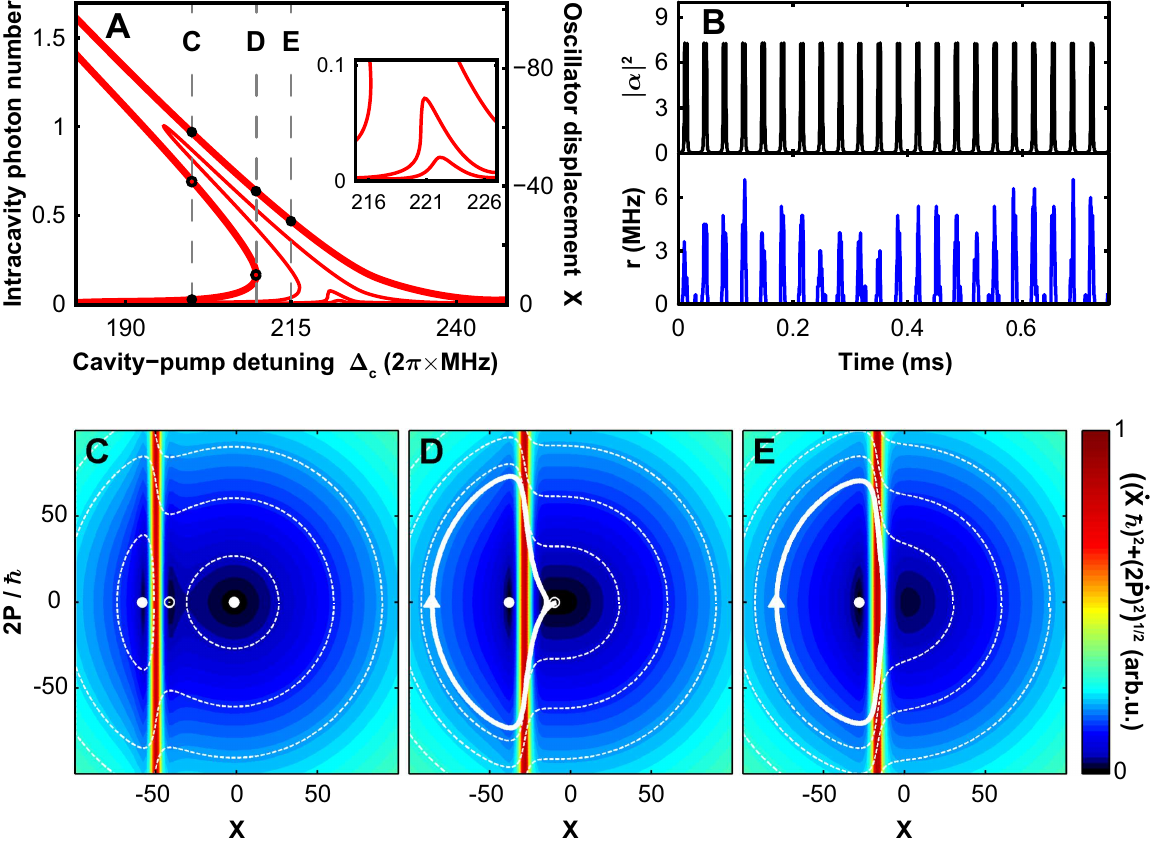}
\caption{Steady state and dynamical behavior of the BEC-cavity system in the two-mode model. (\textbf{A}) Mean intracavity photon number and corresponding oscillator displacement $X$ versus the cavity-pump detuning $\Delta_c$ for the steady state solutions of Eqs.~3. The curves correspond to mean intracavity photon numbers on resonance of $\eta^2/\kappa^2 = 0.02, 0.07, 1$ and $7.3$, and a pump-atom detuning of $\Delta_a = 2\pi\times 32 \, \mathrm{GHz}$. The inset highlights the bistable behavior for pump amplitudes larger than $\eta_\mathrm{cr} \approx 0.27 \kappa$. (\textbf{C-E}) Evolution of the system in the mechanical phase space depicted for three subsequent situations ($\Delta_c = 2\pi\times(200, 209.7, 215)\, \mathrm{MHz}$) corresponding to the markers in A and $\eta^2/\kappa^2 = 7.3$. The stable and unstable steady state configurations are displayed as filled and open circles respectively. Dashed lines show representative evolutions for different starting conditions. Coloring indicates the modulus of the time evolution field $(\dot{X},2\dot{P}/\hbar)$. The solid lines in D and E correspond to the experimental situation in Fig.~2A and show the evolution of the system while scanning $\Delta_c$ at a rate of $2\pi\times 2.9\,\mathrm{MHz}/\mathrm{ms}$ across the resonance with the system initially prepared in the lower stable solution. (\textbf{B}) Intracavity photon number $|\alpha|^2$ and corresponding transmission count rate $r$ (including detection shot noise and averaging over 2 \textmu s) for the system circling along the solid line in E. For integration of the equations of motion a coherent intracavity field $\alpha$ was assumed.}
\end{figure*}

To describe the driven BEC-cavity dynamics we consider a one-dimensional model in which the atomic motion along the cavity axis is quantized. Justified by the large detuning between pump laser frequency and atomic resonance we adiabatically eliminate the internal state dynamics of the atoms. Denoting the creation operator for cavity photons by $\hat{a}^{\dag}$ and the condensate wave function (normalized to the atom number $N$) by $\psi$, the equations of motion for the coupled system read \cite{horak2000,maschler2005}
\begin{eqnarray}
i \hbar \dot{\psi}(x) &=&\Big( \frac{-\hbar^2}{2 m}\frac{d^2}{dx^2} +\langle\hat{a}^{\dag} \hat{a}\rangle \hbar U_0\cos^2(k x) \notag\\
&&+ V_\mathrm{ext}(x) + g_\mathrm{1D}|\psi|^2\Big)\psi(x) \\
i \dot{\hat{a}} &=& -\Big(\Delta_c - U_0\langle \cos^2(k x)\rangle +i \kappa \Big) \hat{a} +i \eta .
\end{eqnarray}
Here, $V_\mathrm{ext}$ denotes the weak external trapping potential for atoms with mass $m$, and $g_\mathrm{1D}$ the effective atom-atom interaction strength integrated along the transverse directions.

Equation 1 describes the condensate dynamics in a dynamic lattice potential. Its depth is determined by the mean intracavity photon number $\langle \hat{a}^\dag \hat{a}\rangle$ which depends in a non-local and non-linear way on the condensate wave function $\psi$ itself. For a single intracavity photon the potential depth is given by the light shift $U_0 = g_0^2/\Delta_a$. The coupling between cavity field and atomic external degrees of freedom is mediated by the spatial overlap $\langle \cos^2(kx)\rangle = \int |\psi(x)|^2 \cos^2(kx)dx$ between atomic density and cavity mode structure, with wavelength $\lambda = 2\pi/k = 780\,\mathrm{nm}$. This mode overlap determines the effective refractive index of the condensate and with it the frequency shift of the empty cavity resonance in Eq.~2. The pump laser which coherently drives the cavity field at a rate $\eta$ is detuned from the empty cavity frequency $\omega_c$ by $\Delta_c = \omega_p - \omega_c$.

The observed BEC-cavity dynamics (Fig.~2) can be described in a homogeneous two-mode model where the macroscopically occupied zero-momentum state is coupled to the symmetric superposition of the $\pm 2\hbar k $ momentum states via absorption and stimulated emission of cavity photons. The corresponding wave function reads
$\psi(x, t) = c_0(t) + c_2(t) \sqrt{2} \cos(2kx)$
with probability amplitudes $c_0$ and $c_2$ fulfilling $|c_0(t)|^2 + |c_2(t)|^2 = N$. The mode overlap is then given by
$\langle \cos^2(kx) \rangle = (N+\sqrt{2} \mathrm{Re}(c_0^* c_2))/2$.
It oscillates under kinetic evolution of $\psi$ at four times the recoil frequency $\omega_\mathrm{rec} = \hbar k^2/(2 m) = 2 \pi \times 3.8 \,\mathrm{kHz}$, with the atom-atom interactions being neglected at this stage. This leads to the natural definition of a harmonic oscillator with displacement $X = 2\sqrt{1/N} \mathrm{Re}(c_0^* c_2)$ in units of the oscillator length, and its conjugate variable $P = \hbar \sqrt{1/N} \mathrm{Im}(c_0^*c_2)$. The equations of motion (Eqs.~1 and 2) then read for $|c_2|^2/|c_0|^2 \ll 1$
\begin{eqnarray}
\ddot{X} + (4\omega_\mathrm{rec})^2 X &=& - \omega_\mathrm{rec} U_0 \sqrt{8 N} \langle \hat{a}^{\dag} \hat{a}\rangle \notag\\
i \dot{\hat{a}} &=& -(\Delta +i \kappa) \hat{a} +i \eta
\end{eqnarray}
and describe a mechanical oscillator coupled via the radiation pressure force to the field of a cavity whose resonance frequency shift $\Delta = \Delta_c - U_0 N/2 - U_0/2 \sqrt{N/2} X$ depends linearly on the oscillator displacement $X$ \cite{kippenberg2007}. The coupling strength between optical and mechanical resonator can be varied via the atom-pump detuning $\Delta_a$ which allows us to experimentally enter the regime of strong coupling.

From this equivalence to cavity opto-mechanics we can anticipate bistable behavior. Indeed, for pump rates larger than a critical value $\eta_\mathrm{cr}$ we find three steady state solutions for the oscillator displacement $X$, with two of them being stable (see Fig.~3A) \cite{meystre1985,dorsel1983,gupta2007}. The system prepared below the resonance will follow the steady state branch until reaching the lower turning point, where a non-steady state dynamics is excited. This dynamics is governed by the time scale of the mechanical motion since the cavity damping is two orders of magnitude faster. Thus we can assume that the cavity field follows the mechanical motion adiabatically and that retardation effects, underlying cooling and amplification, are negligible \cite{kippenberg2007}. Numerical integration of the coupled Eqs.~3 for our experimental parameters results in fully modulated oscillations of the cavity field and cavity output (Fig.~3B), which is in very good agreement with the experimental observations (Fig.~2B).

Further insight is gained by examining the dynamics in the phase space of the mechanical oscillator, spanned by $X$ and $P$ (Fig.~3C-E). Without cavity field the time evolution would simply correspond to a clockwise rotation at $4\omega_\mathrm{rec}$. Yet, when photons enter the cavity the evolution is affected by light forces. This is the case along the vertical resonance line determined by the resonance condition $\Delta = 0$, as shown in Fig.~3C-E (red line).

Initially the condensed atoms are prepared at the stable phase-space point $(X,P)=0$, see Fig.~3C. Increasing the detuning $\Delta_c$ across the resonance renders the system instable and triggers parametrically excited oscillations, as indicated by the solid line in Fig.~3D. The evolution along this path is dominated by the free oscillator dynamics which gets periodically interrupted by the interaction with the cavity light field, Fig.~3D and E. This behavior is closely related to the matter-wave dynamics of a kicked rotor which is operated at an antiresonance where the accumulated phase factor between two kicks inhibits occupation of higher momentum modes \cite{moore1995}.

\begin{figure}
\centering
\includegraphics{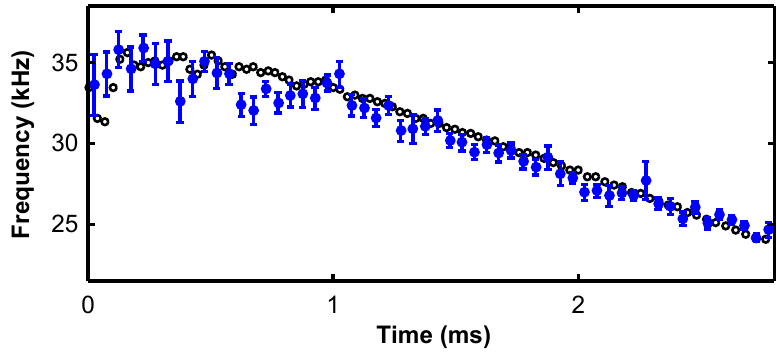}
\caption{Oscillation frequency while scanning over the resonance. The frequency within time bins of 50 \textmu s was obtained from a peak-detection routine applied to the cavity transmission data averaged over 10 \textmu s. The data (filled circles) is an average over 23 traces referenced to the start of the oscillations. The error bars indicate the standard deviation of the mean. Open circles show the result of a numerical integration of the 1D system taking atomic interactions and external trapping into account (Eqs.~1 and 2)\cite{seesom}. The mean intracavity photon number on resonance was $3.6\pm0.9$. To fit the slope of the data the effect of a dynamically induced atom loss during the time of oscillations of $1.5\cdot 10^3/\mathrm{ms}$ was added to the experimental frequency chirp of $\dot{\Delta}_c = 2.9\,\mathrm{MHz}/\mathrm{ms}$. The background rate of atom loss was measured to be $45/\mathrm{ms}$, and an atom number of $(116\pm18)\times 10^3$ was deduced from absorption images taken after the oscillations.}
\end{figure}

The frequency of these oscillations decreases continuously over observation time (see Fig.~4). This is expected when actively scanning the cavity-pump detuning $\Delta_c$ which shifts the resonance line in the phase space diagram and leads to an adiabatic change of the system's circling path (compare Fig.~3D and E).

A precise quantitative understanding of the observed frequency and its decrease is obtained when taking atom-atom interactions, the external trapping potential and atom losses into account. The atom-atom interactions result in a shift of the bare oscillation frequency $4 \omega_\mathrm{rec} = 2\pi \times 15.1\,\mathrm{kHz}$ by the mean field energy, which in the Thomas-Fermi limit equals 4/7 times the chemical potential $ \mu = 2\pi \times 2.4 \,\mathrm{kHz}$ \cite{stenger1999}. The trapping potential gives rise to a Fourier-limited broadening of the initial momentum distribution and accordingly introduces a damping of the free running oscillator dynamics. This suppresses a double peak structure in the transmitted light which would be expected at the onset of oscillations for the homogeneous two-mode model (see Fig.~3D). An enhanced atom loss during the oscillations accelerates the observed frequency shift by a factor of 2. The numerical integration of the full 1D model (Eqs.~1 and 2) yields very good agreement with our data (Fig.~4).

The quantitative agreement between experiment and semi-classical theory, together with the observation of very narrow peaks in the fully modulated cavity transmission, indicates that our system is well localized in the phase space of the mechanical oscillator. Using a second quantized picture where the Bose-Einstein condensate acts as the vacuum state of the mechanical oscillator mode, we have estimated the expectation value for thermal excitations in this mode. It is found to be below 0.01 for a realistic condensate fraction of 90\% \cite{seesom}. This extremely pure preparation of the ground state of a mesoscopic mechanical oscillator is possible since the cavity couples only to one specific excitation mode. Due to the high finesse of the cavity a single coherent mechanical excitation leads to a detectable shift of the cavity resonance by $0.7\kappa$. Entering this strongly-coupled quantum regime of cavity opto-mechanics promises to be ideal for testing fundamental questions of quantum mechanics \cite{mancini1997,marshall2003,zhang2003}.

From the perspective of quantum many-body physics we have investigated a Bose gas with weak local interactions subject to non-local interactions mediated by the cavity field. Experimentally it should also be possible to enter the strongly-correlated regime where local interactions dominate over the kinetic energy. In this case the non-local coupling is predicted to give rise to novel quantum phases \cite{larson2008,maschler2008,nagy2008}.

\section*{Materials and Methods}
\subsection*{Quantum mechanical mapping to cavity opto-mechanics}
To obtain a fully quantized description of the coupled BEC-cavity system we start from its Hamiltonian in second quantized form. Its one dimensional version reads after elimination of the internal excited-state dynamics and in a frame rotating at the pump laser frequency $\omega_p$ \cite{horak2000,maschler2005}
\begin{eqnarray}
\hat{H} &=&
\int \hat{\Psi}^\dag(x)\Big(\frac{-\hbar^2}{2m}\frac{d^2}{dx^2}+V_\mathrm{ext}(x)\notag \\
&&\qquad \qquad \ \ +\hbar U_0 \cos^2(k x) \hat{a}^\dag \hat{a}\Big)\hat{\Psi}(x)\,dx \notag \\
&&+\hat{H}_{A-A} - \hbar \Delta_{c} \hat{a}^\dag \hat{a} - i \hbar \eta (\hat{a}-\hat{a}^\dag) + \hat{H}_\kappa.\notag
\end{eqnarray}
Here, $\hat{\Psi}^{\dag}$ denotes the creation operator of atoms with mass $m$, and $\hat{a}^{\dag}$ that of cavity photons with frequency $\omega_c$, wave vector $k = 2\pi/\lambda$ and mode function $\cos(kx)$. The maximum light shift which an atom experiences in the cavity mode is given by $U_0 = g_0^2/\Delta_a$ with the atom-photon coupling constant $g_0$. The pump laser frequency is detuned from the empty cavity resonance frequency $\omega_c$ and the atomic transition frequency $\omega_a$ by $\Delta_c = \omega_p - \omega_c$ and $\Delta_a = \omega_p - \omega_a$ respectively. Decay of cavity photons at a rate $\kappa$ is accounted for by the term $\hat{H}_\kappa$.

In case of weak atom-atom interactions $\hat{H}_{A-A}$ and a shallow external trapping potential $V_\mathrm{ext}$ the BEC-cavity system can be mapped onto the generic Hamiltonian of cavity opto-mechanics \cite{ludwig2008}. To this end we expand $\hat{\Psi}(x)$ into the two spatial modes $\phi_0(x) = 1$ and $\phi_2(x) = \sqrt{2} \cos(2kx)$ which dominantly contribute to the BEC-cavity dynamics. The corresponding bosonic annihilation operators are denoted by $\hat{c}_0$ and $\hat{c}_2$. Applying the Bogoliubov approximation $\hat{c}_0 = \sqrt{N}$ and taking $\langle \hat{c}^\dag_2 \hat{c}_2\rangle \ll N$ into account we get
\begin{eqnarray}
\hat{H} &=& 4 \hbar \omega_\mathrm{rec} \hat{c}_2^\dag \hat{c}_2+ \hbar\Big({-\tilde{\Delta}_c} + g(\hat{c}_2 + \hat{c}_2^\dag) \Big)\hat{a}^\dag \hat{a} \notag \\
&&- i \hbar \eta (\hat{a}-\hat{a}^\dag) + \hat{H}_\kappa\notag.
 \end{eqnarray}
Due to the presence of the atoms we obtain a shifted cavity-pump detuning of $\tilde{\Delta}_c = \Delta_c - \frac{1}{2} U_0 N$. The matter-wave mode $\phi_2$ plays the role of a quantum-mechanical oscillator with its oscillation frequency $4 \omega_\mathrm{rec}$, determined by the kinetic energy of this matter-wave mode. The oscillator is coupled to the cavity field with a collectively enhanced coupling strength $g = U_0/2 \sqrt{N/2}$. To which extend quantum fluctuations play a role in the system is determined by the ratio $g/\kappa$ \cite{ludwig2008}. For our experimental parameters we obtain $g/\kappa = 0.3\,(0.6)$ for the $\sigma^- (\sigma^+)$ transition.  This coupling strength can be tuned via the pump-atom detuning $\Delta_a$ and the atom number $N$ which allows us to experimentally enter the strong coupling regime of cavity opto-mechanics.

From our coupling strength we can deduce the effective oscillator mass $m_\mathrm{eff}$ in the cavity opto-mechanical model system \cite{ludwig2008,kippenberg2007}. There the radiation pressure coupling strength is given by $g = \omega_c a_\mathrm{ho}/L$ where $L$ denotes the length of the cavity and $a_\mathrm{ho} = \sqrt{\hbar/(2 m_\mathrm{eff} \omega_m)}$ the harmonic oscillator length with oscillator frequency $\omega_m=4\omega_\mathrm{rec}$. With this we obtain an effective mass of $m_\mathrm{eff} = 0.01\,\mathrm{ng}$.

\subsection*{Ground state preparation}
We calculate the broadening of the overlap operator $\hat{u} = \int dx \cos^2(k x)\hat{\Psi}^\dag(x) \hat{\Psi}(x)$ caused by thermal depletion of the condensate. To estimate the number of thermal excitations in the mechanical oscillator mode we compare this broadening with the zero-point fluctuations of $\hat{u}$ for the mechanical oscillator being in its ground state. We restrict the discussion to the direction along the cavity axis since only density fluctuations along this axis contribute. The atomic annihilation operator $\hat{\Psi}(x)$ is split into a condensate part and a thermal part
\begin{equation}
\hat{\Psi}(x) = \varphi_0(x) \sqrt{N_0} + \sum_{i\neq 0}\varphi_i(x)\hat{a}_i.\notag
\end{equation}
Here $\varphi_0$ denotes the condensate wave function obtained in the Thomas-Fermi approximation for $N_0$ condensed atoms, and $\varphi_i$  the harmonic oscillator eigenfunctions in the external trapping potential. For simplicity we neglect the effect of atom-atom interactions on the thermal atoms. We evaluate the variance of $\hat{u}$ in a thermal state of temperature $T$ where the uncondensed atoms $N_T = N - N_0$ are distributed over the excited state levels according to a Bose distribution with chemical potential $\mu = 0$. The total number of atoms $N$ is kept fixed and the condensate fraction is given by $N_0/N = 1-(T/T_c)^3$ with the critical temperature $T_c$ calculated using the external trapping frequencies. Beside autocorrelations in the density fluctuations we find for the variance $\Delta u^2$
\begin{eqnarray}
\Delta u^2 \equiv \langle \hat{u}^2\rangle-\langle \hat{u} \rangle^2 &=& 2 N_0 \sum_{i\neq 0} M_{0i}^2 \langle\hat{a}_i^\dag\hat{a}_i\rangle \notag \\
&&+ \sum_{i, j\neq 0}M_{ij}^2\langle\hat{a}_i^\dag\hat{a}_i\rangle\langle\hat{a}_j^\dag\hat{a}_j\rangle
\end{eqnarray}
where we introduced the matrix elements $M_{ij} = 0.5\langle \varphi_i|\cos(2kx)|\varphi_j\rangle$. The first term on the right hand side of Eq.~4 originates from interference between thermal atoms and the condensate. For $N_T<N_0$ this contribution by far exceeds the second term which corresponds to purely thermal density fluctuations at a wave vector $2 k$. Comparing $\Delta u^2$ with the zero-point fluctuations of the overlap $\langle 0|(\hat{u}-N/2)^2 |0\rangle = N/8$ at $T=0$ gives an estimate for the number $n_T = 8\frac{\Delta u^2}{N}$ of thermal excitations in the mechanical oscillator mode. Numerically we find a thermal occupation $n_T$ below 0.01 for a condensate fraction of 90\%.

\subsection*{Numerical integration of the semi-classical 1D model}
The integration of the semi-classical Eqs.~1 and 2 including the external trapping potential and atom-atom interactions is done on a spatial grid of size $13\lambda \approx 10$\,\textmu m with 10 points per $\lambda/2$. The cavity damping rate is two orders of magnitude larger than $4\omega_\mathrm{rec}$ which characterizes the time scale of atomic motion. Therefore the intracavity field dynamics is adiabatically eliminated and described by the coherent state variable $\alpha(t)$. Quantum fluctuations of the cavity light field are not taken into account. The initial ground state solution for $\Delta_c$ below the optical resonance is obtained numerically using imaginary time propagation. Subsequently, this solution is propagated in real time with times steps of $dt = 100\,\mathrm{ns}$ while scanning $\Delta_c$ across the resonance.

\section{Acknowledgments}
\begin{acknowledgments}
We would like to thank Kristian Baumann, Peter Domokos, Christine Guerlin, Igor Mekhov, Helmut Ritsch and Andr\'{a}s Vukics for stimulating discussions and acknowledge funding by the SCALA Integrated Project (EU) and QSIT (ETH).
\end{acknowledgments}

\end{document}